\documentclass[12pt,twoside]{article}
\usepackage[mathscr]{eucal}
\usepackage{amsmath,amsfonts,amssymb,amsthm}
\usepackage{hyperref}
\usepackage[matrix,arrow]{xy}
\usepackage{times}
\usepackage{pdfsync}
\usepackage{graphicx}
\usepackage{epstopdf}
\usepackage{multibox}
\usepackage{dcolumn}

\def\d_V{\text{d}_V}

\def\d_Vphi{\text{d}_V\hspace{-0.06em}\phi}
\def\d_Vphibar{\text{d}_V\hspace{-0.06em}\bar\phi}
\def\d_Vxi{\text{d}_V\hspace{-0.06em}\xi}

\def\be{\begin{eqnarray}}
\def\ee{\end{eqnarray}}
\def\beann{\begin{eqnarray*}}
\def\eeann{\end{eqnarray*}}
\def\beq{\begin{equation}}
\def\eeq{\end{equation}}
\def\ba{\begin{array}}
\def\ea{\end{array}}
\def\ben{\begin{enumerate}}
\def\een{\end{enumerate}}
\def\bea{\begin{eqnarray}}
\def\eea{\end{eqnarray}}

\def\5{\bar }
\def\6{\partial }
\def\7{\hat }
\def\4{\tilde }
\advance\textheight\topskip
\renewcommand{\tilde}{\widetilde}
\renewcommand{\hat}{\widehat}

\renewcommand{\d}{\partial}

\newcommand{\binner}[2]{%
  {\langle}\kern-4.15pt{\langle}#1{,}\,#2{\rangle}\kern-4.15pt{\rangle}}

\newcommand{\ffrac}[2]{\raisebox{.5pt}%
  {\footnotesize$\displaystyle\frac{#1}{#2}$}\kern1pt}

\numberwithin{equation}{section} \makeatletter
\@addtoreset{equation}{section}

\begin{document}

\def\mytitle{The Coulomb solution as a coherent state of unphysical photons}

\pagestyle{myheadings} \markboth{\textsc{\small Glenn Barnich}}{%
  \textsc{\small Quantum Coulomb solution}} \addtolength{\headsep}{4pt}

\begin{flushright}\small
ULB-TH/10-01
\end{flushright}

\begin{centering}

  \vspace{1cm}

  \textbf{\Large{\mytitle}}

  \vspace{1cm}

  {\large Glenn Barnich$^{a}$}

\vspace{1cm}

\begin{minipage}{.9\textwidth}\small \it \begin{center}
   Physique Th\'eorique et Math\'ematique \\ Universit\'e Libre de
   Bruxelles\\ and \\ International Solvay Institutes \\ Campus
   Plaine C.P. 231, B-1050 Bruxelles, Belgium \end{center}
\end{minipage}

\end{centering}

\vspace{.5cm}

\begin{center}
  \begin{minipage}{.9\textwidth}
    \textsc{Abstract}. In the context of the problem of what
    micro-states are responsible for the entropy of black holes, we
    consider as a physical toy model the electromagnetic Coulomb
    solution. By quantizing the electromagnetic field in the presence
    of an external source of charge $Q$, the quantum state
    corresponding to the Coulomb solution is identified as a coherent
    state of longitudinal and temporal photons in a Hilbert space with
    negative norm states.
  \end{minipage}
\end{center}

\vspace{.5cm}

\begin{minipage}{.9\textwidth}

\end{minipage}

\vfill

\noindent
\mbox{}
\raisebox{-3\baselineskip}{%
  \parbox{\textwidth}{\mbox{}\hrulefill\\[-4pt]}}
{\scriptsize$^a$Research Director of the Fund for
  Scientific Research - FNRS (Belgium). E-mail: gbarnich@ulb.ac.be}

\thispagestyle{empty}
\newpage


\newpage

\section{Introduction}

A major challenge in contemporary theoretical physics is a better
understanding of the microscopic origin of the Bekenstein-Hawking
black hole entropy \cite{Bekenstein:1973ur,Hawking:1974sw}. There have
been many speculations that the degrees of freedom responsible for
this entropy might have something to do with gauge degrees of freedom
(see e.g.~\cite{thooftsolvay}) that cease to be pure gauge because of
non trivial boundary conditions. This is borne out by the existence of
a black hole solution in 2+1 dimensional gravity with a negative
cosmological constant \cite{Banados:1992wn,Banados:1993gq}, despite
the fact that there are no physical gravitons. Because of the latter
feature, the problem of the entropy of the BTZ black hole is
technically simpler than the one of its higher dimensional cousins,
and has been analyzed extensively (see
e.g.~\cite{Carlip:1994gy,Carlip:1998qw,Strominger:1998eq}).

From a mathematical perspective, mass and angular momentum in general
relativity are akin to electric charge in pure electromagnetism: these
observables are described by surface charges and are related to
parameters of vanishing gauge transformations, Killing vectors of a
suitable background in the former case and a constant gauge parameter
in the latter \cite{Abbott:1981ff,Abbott:1982jh}. From this point of
view, the analogue of a black hole of mass $M$ is the Coulomb solution
with charge $Q$. Before turning to black holes, it is thus useful to
understand the micro-states corresponding to the Coulomb solution.

The computation presented below is a straightforward exercise whose
result is known in some form or the other (see e.g.~section 18 of
\cite{Papanicolaou:1976sv}) and that can easily be generalized to the
coupling of the electromagnetic field with an arbitrary conserved
current. Nevertheless, it seems worthwhile to provide a complete and
self-contained Hamiltonian derivation, as it sheds some light on the
black hole problem.

\section{Quantum Coulomb solution as a coherent state}

By Coulomb solution, we mean the solution to Maxwell's equations with
zero magnetic and transverse electric fields that is produced by a
static point-particle charge. For simplicity, we consider the case
where the world-line of the point-particle coincides with the
time-axis of Minkowski space-time. The associated action principle is
given by
\begin{eqnarray}
S^Q=\int d^4x\ [-\frac{1}{4}F_{\mu\nu}F^{\mu\nu}-j^\mu
A_\mu],\label{eq:35}
\end{eqnarray}
with $j^\mu=\delta^\mu_0Q\delta^3(\vec x)$.  

BRST quantization of this system is well-known. We follow chapter 19
of \cite{Henneaux:1992ig}\footnote{Note that (19.33a), (19.33d) should
  be multiplied by $\sqrt 2$ and (19.33b), (19.33c) divided by $\sqrt
  2$ in order to yield the BRST charge (19.34).}, where the case
$j^\mu=0$ has been considered in great detail.  In the presence of the
source, the secondary constraint, Gauss's law, becomes
$\phi^Q_2\equiv-\pi^i,_i+j^0= 0$ and the BRST charge is modified to
\begin{eqnarray}
  \Omega^Q=\int d^3k\ 
 c^*(\vec k) [ a(\vec k)-q(\vec k)]+[ a^*(\vec
k)-q(\vec k)] c(\vec k),\label{eq:36}
\end{eqnarray}
where $a(\vec k)= a_3(\vec k)+ a_0(\vec k)$, with $a_3(\vec
k),a_0(\vec k)$ the oscillators associated with longitudinal and
temporal photons respectively and $q(\vec
k)=\frac{Q}{(2\pi)^{3/2}\sqrt 2 k^{3/2}}$.

The BRST operator $\hat\Omega^Q$ does not any longer annihilate the
standard vacuum. In order to have the usual quartet mechanism
\cite{Kugo:1979gm} at work, one can define the shifted oscillators
$a^Q(\vec k)=a(\vec k)-q(\vec k)$, $a^{*Q}=a^*(\vec k)-q(\vec k)$
without affecting the commutation relations and define the appropriate
vacuum through
\begin{eqnarray}
\hat a^Q(\vec k)|0\rangle^Q=0,\label{eq:37}
\end{eqnarray}
with $|0\rangle^Q$ being annihilated by all the other destruction
operators. In terms of the standard vacuum, the new vacuum appears as a
coherent state made out of temporal and longitudinal photons, 
\begin{eqnarray}
|0\rangle^Q=\prod_{\vec k}\exp {q(\vec k)\hat b^\dagger(\vec
  k)}|0\rangle
=\exp{\int
  d^3k\ q(\vec k)\hat b^\dagger(\vec k)}\ |0\rangle ,\label{eq:38}
\end{eqnarray}
with $b(\vec k)=\frac{1}{2}(a_3(\vec k)-a_0(\vec k))$ and $[\hat
a(\vec k),{\hat b}^\dagger(\vec k^\prime)]=\delta^3(\vec k-\vec
k^\prime)$.  Because ${\hat b}^\dagger(\vec k)$ are null oscillators,
${}^Q\langle 0|0\rangle^Q=\langle 0|0\rangle =1$. As we are now going
to show, it is the state
$|0\rangle^Q$ that corresponds to the Coulomb solution.

Indeed, the expectation value of $\hat A_\mu(\vec x)$ in the new
vacuum is
\begin{eqnarray} 
{}^Q\langle 0|\hat A_\mu(\vec x)|0\rangle^Q=
\delta_\mu^0\frac{1}{(2\pi)^{3/2}}\int 
d^3k\ \frac{Q}{(2\pi)^{3/2}2 k^2}\exp{i\vec k\vec x}  
  =\delta_\mu^0\frac{Q}{8\pi r}.\label{eq:39}
\end{eqnarray}
With $A_i=0$, the classical equations of motion are solved by 
\begin{eqnarray}
A_\mu(\vec
x)=\delta_\mu^0\frac{Q}{4\pi r},\label{eq:41}
\end{eqnarray}
which differs from \eqref{eq:39} by a
factor $2$. There is however no contradiction as 
\begin{eqnarray}
{}^Q\langle 0|\partial_i\hat A^i(x)|0\rangle^Q\neq 0.\label{eq:42}
\end{eqnarray}
even though ${}^Q\langle 0|\hat A_i(x)|0\rangle^Q= 0$ and thus
$\partial_i {}^Q\langle 0|\hat A^i(x)|0\rangle^Q= 0$ since the
polarization vector $e^3_i(\vec k)$ is odd under $\vec k\rightarrow
-\vec k$, while $k^ie^3_i(\vec k)$ is even.

The expectation values of the gauge invariant electric
and magnetic fields in the new vacuum agree with their classical
analogs:
\begin{eqnarray}
{}^Q\langle 0|\hat
\pi^i(x) |0\rangle^Q&=& {}^Q\langle 0|-\frac{i}{(2\pi)^{3/2}}\int
d^3k\sqrt{\frac{k}{2}}\ [a\, e_3^i\,\exp{i\vec k\vec
  x}-a^*\,e^i_3\,\exp{-i\vec k\vec x}]
|0\rangle^Q\nonumber\\ &=&-\frac{2i}{(2\pi)^{3/2}}\int
d^3k\sqrt{\frac{k}{2}}\,q(\vec k) \frac{k_i}{k }\exp{i\vec
  k\vec x}\nonumber\\ &=&-\frac{Q}{(2\pi)^{3}}\frac{\partial}{\partial x^i}\int
d^3k\ \frac{1}{k^2} \exp{i\vec k\vec
  x} = \frac{Qx^i}{4\pi r^3},\label{eq:44}
\end{eqnarray}
while 
\begin{eqnarray}
{}^Q\langle 0|\vec\nabla\times\hat{\vec A}|0\rangle^Q=0 \label{eq:45}
\end{eqnarray}
because $\epsilon^{ijl}k_j e^3_l(\vec k)=0$. 

\section{Discussion}
\label{sec:discussion}

Since it is built out of longitudinal and temporal photons,
understanding the Coulomb solution quantum mechanically is thus only
possible if one quantizes all polarizations in a Hilbert space with
negative norm states and not if one reduces to physical degrees of
freedom around the new vacuum before quantization. The advantage as
compared to the $A_0=0$ gauge advocated for instance in
\cite{Gervais:1978kn} is that a charge eigenstate now appears as a
displaced vacuum state. This generalizes to unphysical photons the
corresponding property of physical photons when coupled to a classical
external vector current density \cite{Glauber:1963tx}.

\section*{Acknowledgments}

The author wants to thank M.~Ba\~nados, J.~Evslin, A.~Gomberoff,
C.~Krishnan, C.~Maccaferri, C.~Schomblond and S.~Theisen for useful
discussions. This work is supported in part by the Fund for Scientific
Research-FNRS (Belgium), by the Belgian Federal Science Policy Office
through the Interuniversity Attraction Pole P6/11, by IISN-Belgium, 
by Fondecyt projects No.~1085322 and No.~1090753 and by the
Perimeter Institute for Theoretical Physics.


\def\cprime{$'$}
\providecommand{\href}[2]{#2}\begingroup\raggedright\endgroup

\end{document}